\documentclass{INTERSPEECH2023}

\interspeechcameraready

\usepackage[dvipsnames]{xcolor}
\usepackage{multirow}

\title{Deep Speech Synthesis from MRI-Based Articulatory Representations}
\name{Peter Wu$^1$, Tingle Li$^1$,  Yijing Lu$^3$,  Yubin Zhang$^3$, Jiachen Lian$^1$, Alan W Black$^2$, Louis Goldstein$^3$, Shinji Watanabe$^2$, Gopala K. Anumanchipalli$^1$}
\address{
$^1$University of California, Berkeley, United States\\
$^2$Carnegie Mellon University, United States\\
$^3$University of Southern California, United States}
\email{peterw1@berkeley.edu}

\begin{document}

\maketitle
 
\begin{abstract}


In this paper, we study articulatory synthesis, a speech synthesis method using human vocal tract information that offers a way to develop efficient, generalizable and interpretable synthesizers. While recent advances have enabled intelligible articulatory synthesis using electromagnetic articulography (EMA), these methods lack critical articulatory information like excitation and nasality, limiting generalization capabilities. To bridge this gap, we propose an alternative MRI-based feature set that covers a much more extensive articulatory space than EMA. We also introduce normalization and denoising procedures to enhance the generalizability of deep learning methods trained on MRI data. Moreover, we propose an MRI-to-speech model that improves both computational efficiency and speech fidelity. Finally, through a series of ablations, we show that the proposed MRI representation is more comprehensive than EMA and identify the most suitable MRI feature subset for articulatory synthesis.

\end{abstract}
\noindent\textbf{Index Terms}: speech synthesis, articulatory synthesis

\section{Introduction}
\label{sec:intro}

Deep speech synthesis technology has made significant advancements in recent years, leading to high-performing models for tasks such as text-to-speech \cite{hayashi2021espnet2, wu2021crosslingual, jets22lim}, voice conversion \cite{polyak21facebookresynthesis, lakhotia2021generative}, and speech translation \cite{chen2022wavlm, deng2022blockwise}. However, the development of brain-to-speech devices \cite{anumanchipalli2019speech, metzger2022generalizable} still poses significant challenges, requiring faster and more data-efficient models. Articulatory synthesis \cite{fant1991articulatorysynthesis, rubin1981articulatorysynthesis, scully1990articulatorysynthesis, wu2022deep, yu2021reconstructing, beguvs2022articulation} offers a potential solution by synthesizing speech from a compact, smooth, and interpretable articulatory space \cite{toda2004acoustic, wu2023speaker, cho2022evidence, lim2021multispeaker, jiachen2022gesture, lian2022gesture2}.

Electromagnetic articulography (EMA) is a commonly used articulatory representation \cite{wu2022deep}, but it only contains 6 x-y points, making it challenging to comprehensively capture articulatory movements. Real-time magnetic resonance imaging (MRI) is a state-of-the-art tool that captures dynamic information about vocal tract movements and shaping during human speech production, offering a feature-rich alternative to EMA. It contains hundreds of x-y points, including positional information for the hard palate, pharynx, epiglottis, velum, and larynx, all of which are important for speech production but not directly described in raw EMA data. Moreover, recent advances in image acquisition and reconstruction techniques have enabled sufficient temporal and spatial resolutions ({\em e.g.}, 12ms and 2.4 × 2.4 mm$^2$) that allow researchers to study the intricate and dynamic interactions during speech production \cite{lingala2016recommendations, lingala2017fast}.

However, since MRI dataset participants speak from inside a tube-shaped MRI machine, there is a noticeable amount of reverberation in the collected utterances, resulting in an unsatisfactory performance on MRI-to-speech synthesis. To overcome this problem, we enhance the utterances and propose a generative adversarial network (GAN) based method that directly synthesizes waveform from articulatory features, which produces noticeably more intelligible speech than the baselines.

We summarize our contributions as follows: 
\begin{itemize}[leftmargin=8.5mm]
    \item We propose a novel MRI-based representation for articulatory synthesis, along with effective preprocessing strategies for such data.
    \item We demonstrate that our proposed model outperforms baselines across several evaluation metrics.
    \item We quantitatively and qualitatively identify the advantages of MRI over EMA and the most important MRI features for articulatory synthesis.
\end{itemize}

Code and additional related information will be available at \href{https://github.com/articulatory/articulatory}{https://github.com/articulatory/articulatory}.

\section{MRI Dataset}
\label{sec:dataset}

We utilize the real-time MRI and its corresponding audio recordings of one native American English speaker (female, 25-year-old), with a total speech duration of approximately 17 minutes and a sampling rate of 20 kHz, which are acquired from a publicly available multispeaker MRI dataset \cite{lim2021multispeaker}. This dataset includes midsagittal real-time MRI videos with a spatial resolution of 2.4 × 2.4 mm$^2$ (84 × 84 pixels) and a temporal resolution of 12-ms (83 frames per second), capturing the vocal tract movements during the production of a comprehensive set of scripted and spontaneous speech materials.

\begin{figure}[t]
\centering
  \includegraphics[width=60mm]{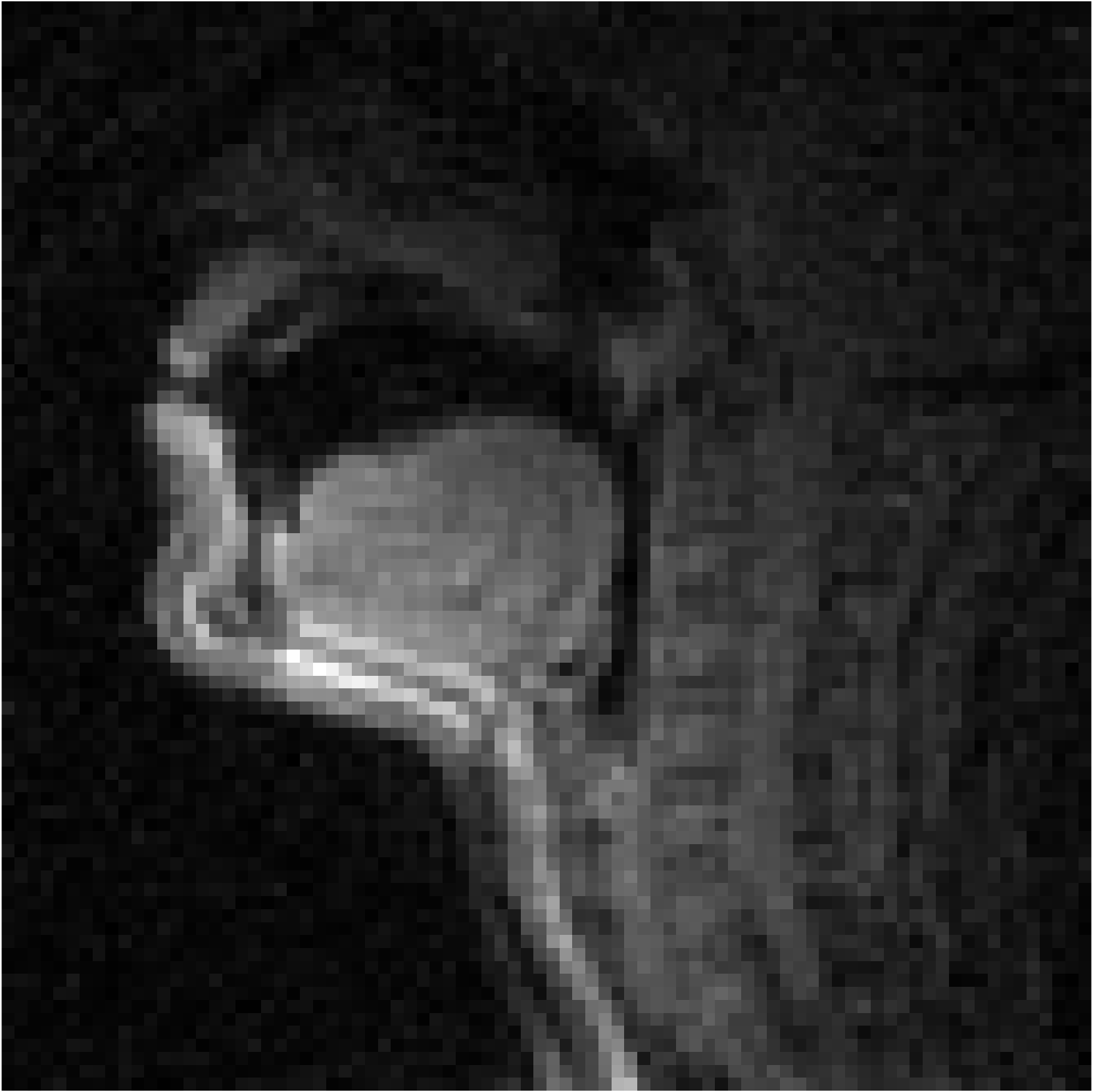}
  \caption{One MRI frame during the utterance ``apa".}
  \label{fig:RAWmri}
\end{figure}
\begin{figure}[t]
\centering
  \includegraphics[width=60mm]{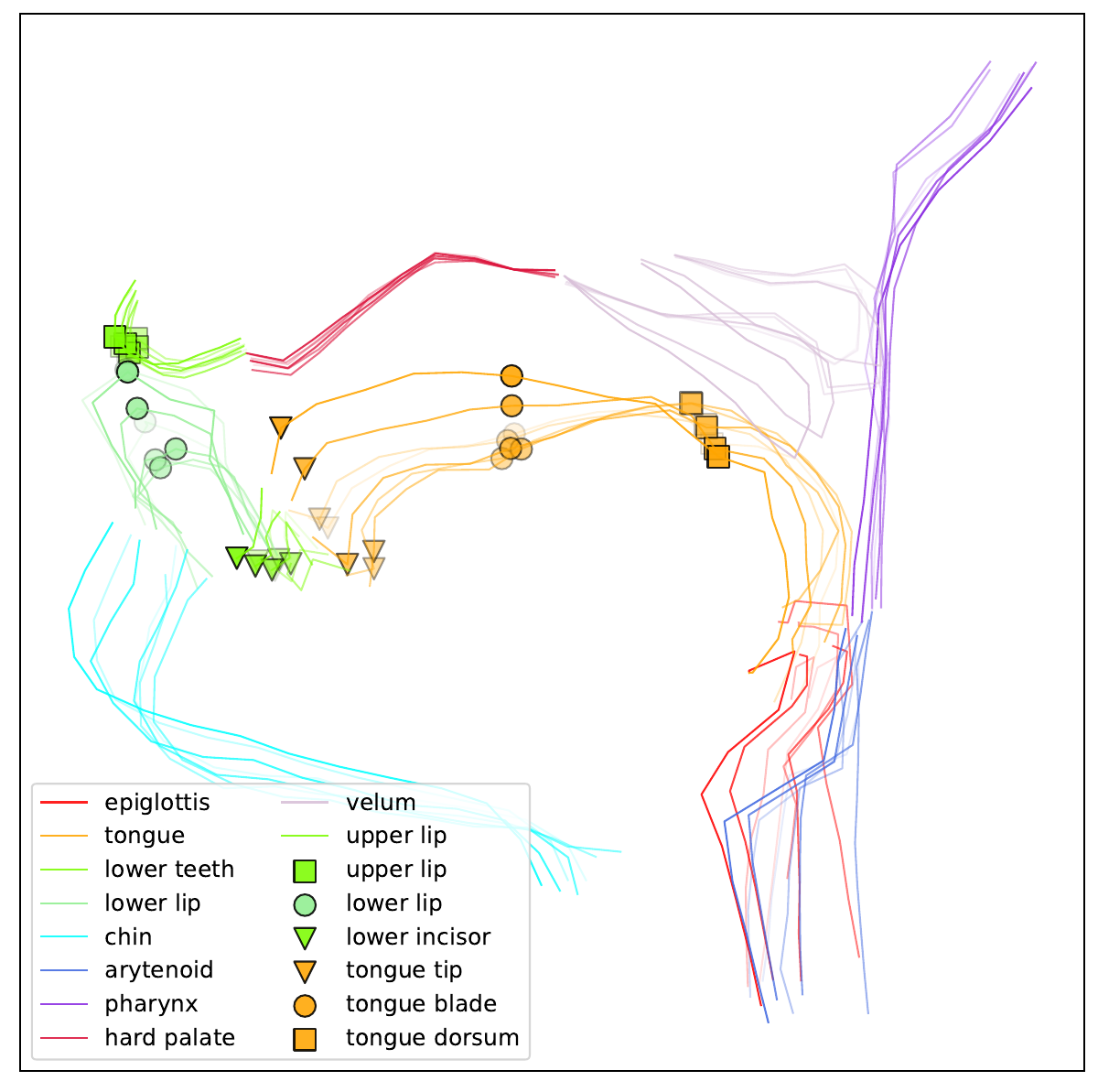}
  \caption{Extracted MRI features for the utterance ``apa". Lighter is earlier in time. The labeled points are the estimated EMA features (Sec. \ref{sec:dataset}).}
  \label{fig:mri_apa}
\end{figure}
\begin{figure}[t]
\centering
  \includegraphics[width=60mm]{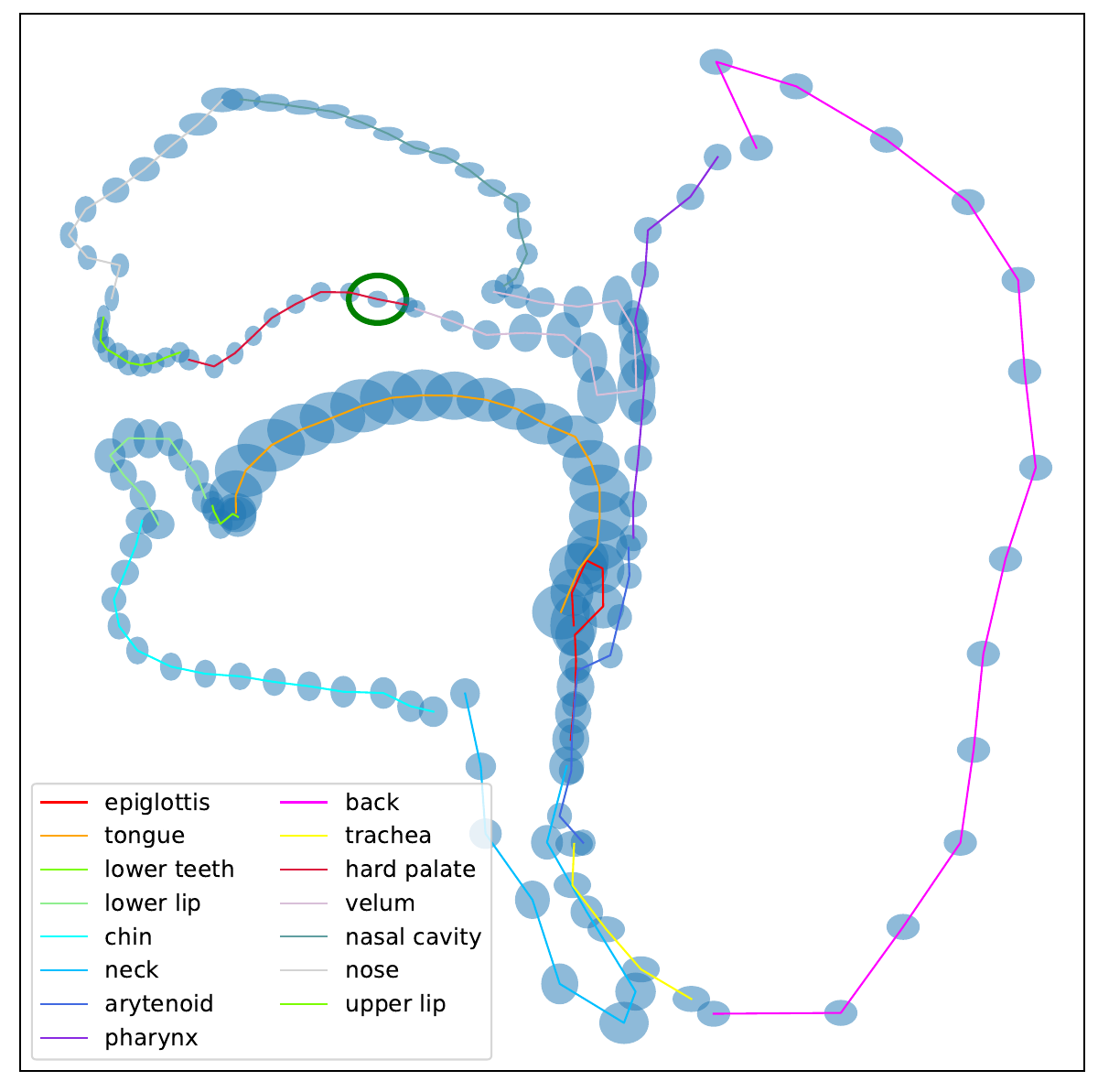}
  \caption{Standard deviation of each MRI feature (Sec. \ref{sec:dataset}).}
  \label{fig:mri_stdev}
\end{figure}

To prepare the MRI data for our model, we use a semi-automatic method \cite{bresch2008region} to track the contours of vocal tract air-tissue boundaries in each raw MRI frame (Figure \ref{fig:RAWmri}) and segmented the contours into anatomical components, as shown in Figure \ref{fig:mri_apa} and Figure \ref{fig:mri_stdev}. To mitigate the problem of overfitting, we pruned the MRI feature set by discarding segments that did not contribute much to understanding how speech production varies across utterances. Figure \ref{fig:mri_stdev} presents the full set of segments, while Figure \ref{fig:mri_apa} shows the reduced set. The original set comprises 170 x-y coordinates, whereas the reduced set contains only 115. We then concatenate and flatten the 115 x-y coordinates into a 230-dimensional vector, which we used as input for our MRI-to-speech synthesis task.


Since the raw MRI data is composed of long utterances with lots of silences, we first segment the utterances into sentence-long pieces. Then we employ a pre-trained BERT-based model\footnote{\url{https://huggingface.co/felflare/bert-restore-punctuation}} to estimate sentence boundaries, and align the audio recordings as well as the orthographic and phonological transcriptions using Montreal force aligner \cite{McAuliffe2017MontrealFA}. The resulting alignments are manually calibrated by professional phoneticians. By utilizing the estimated sentence boundaries and word alignments, we split the audio recordings and MRI data into 236 utterances, totaling 11 minutes. Finally, these utterances are randomly split into a 0.85-0.05-0.10 train-val-test split, resulting in 200, 11, and 25 utterances in the train, val, and test sets, respectively.

Furthermore, the head location is not fixed within the MRI data, which can negatively impact the ability of such models to generalize to unseen positions. Thus, we adopt a centering approach that centers each frame around a relatively fixed point to improve generalizability. Specifically, we calculate the standard deviation $(\sigma_x, \sigma_y)$ of each of the 170 points across the training set and center every frame at the point with the lowest $\sqrt{\sigma_x^2 +\sigma_y^2}$. This center point is located on the hard palate, circled in green in Figure \ref{fig:mri_stdev}. We note that the standard deviation results reflect human speech production behavior, as the hard palate is relatively still across utterances whereas the tongue varies noticeably, highlighting the interpretability of our MRI-based articulatory features. Another preprocessing step that we found useful was denoising, which we detail in Section \ref{sec:enhance}.

\section{Models}
\label{sec:models}

\subsection{Intermediate-Representation Baselines}

Currently, a popular speech synthesis approach is to first synthesize an intermediate representation from the input and then map the intermediate representation to the waveform domain \cite{Csapo2020UltrasoundbasedAM, georges2020towards, yu2021reconstructing}. Wu {\em et al.} \cite{wu2022deep} showed that directly synthesizing speech from EMA outperformed a spectrum-intermediate approach in terms of computational efficiency and yielded comparable synthesis quality. Intuitively, omitting the spectrum intermediate reflects how the human speech production process does not perform this intermediate mapping \cite{wu2022deep}. In this work, we also compare using intermediate representations with directly synthesizing from inputs. We observe two popular types of intermediate representations in the literature: (1) spectrums \cite{Csapo2020UltrasoundbasedAM, georges2020towards, yu2021reconstructing}, and (2) deep representations \cite{polyak21facebookresynthesis}. To compare our proposed direct modelling approach in Section \ref{sec:time_domain_hifi} with both intermediate modelling methods, we experiment with Mel spectrogram and HuBERT \cite{hsu2021hubert} intermediates. For the Mel spectrogram calculation, we use size-240 hops, size-1024 FFTs, Hann windows, and 80 Mels. With HuBERT, we use the output of the model's last hidden layer, linearly interpolated to match the MRI input sampling rate. We denote spectrum-intermediate models with ``Spe." and HuBERT ones with ``Hub." in our results below for readability. In our MRI-to-speech task, direct modeling is both more computationally efficient and more high-fidelity than the intermediate approaches, as discussed in Sections \ref{sec:computational_efficiency} and \ref{sec:synthesis_quality}. We detail the model architectures of our intermediate-representation baselines in Section \ref{sec:cnnblstm}.

\subsection{CNN-BiLSTM Baseline}
\label{sec:cnnblstm}

As per Yu {\em et al.} \cite{yu2021reconstructing}, we employ the CNN-BiLSTM architecture as the baseline method. This method involves processing each MRI frame through a sequence of four CNN layers, with two max-pooling layers incorporated in the middle. The extracted features are then aggregated along the time axis and fed to a BiLSTM layer to generate the mel-spectrogram. Since the inputs in our MRI-to-speech task are sequences of vectors rather than the MRI video inputs used in Yu {\em et al.} \cite{yu2021reconstructing}, we use 1D convolutions instead of 2D and 3D. Finally, a neural vocoder is used to reconstruct the waveform signal. For this vocoder, we use HiFi-CAR \cite{wu2022deep}, which outperforms the WaveGlow architecture \cite{prenger2019waveglow} used by Yu {\em et al.} \cite{yu2021reconstructing}. HiFi-CAR is an autoregressive version of the HiFi-GAN convolutional network \cite{kong2020hifigan}, detailed in Section \ref{sec:time_domain_hifi}. It is worth noting that they used the original speech data, without any denoising, resulting in unsatisfactory performance. For a fair comparison, we also train this model using enhanced speech. In our experiments in Sections \ref{sec:computational_efficiency} and \ref{sec:synthesis_quality} below, we refer to this model as CBL for readability.

\subsection{HiFi-CAR Model}
\label{sec:time_domain_hifi}

Similar to the method used by Wu {\em et al.} \cite{wu2022deep}, our model directly synthesizes waveforms from articulatory features without the need for an intermediate representation. Specifically, we build on their HiFi-CAR model, which is a HiFi-GAN convolutional neural network \cite{kong2020hifigan} modified to be autoregressive using the CAR-GAN audio encoder \cite{morrison2022cargan}. To our knowledge, training models to directly synthesize waveforms from MRI data has not yielded successful results previously. However, we observe that this model outperforms our baselines in terms of both computational efficiency and fidelity, as discussed in Sections \ref{sec:computational_efficiency} and \ref{sec:synthesis_quality}. We also use the HiFi-CAR vocoder to map intermediate features to waveforms for our intermediate-representation baselines. For all HiFi-CAR models, we initialize their weights with those of a HiFi-GAN spectrum-to-waveform vocoder pre-trained on LibriTTS.\footnote{\url{https://github.com/kan-bayashi/ParallelWaveGAN}} We note that this initialization approach noticeably improves performance compared to Wu {\em et al.} \cite{wu2022deep}. Further modeling details can be found in the accompanying codebase.

\subsection{Speech Enhancement Model}
\label{sec:enhance}

The currently available dataset \cite{lim2021multispeaker} suffers from poor quality due to significant reverberation and noise, which poses a significant challenge for accurate modeling of the relationship between MRI and speech. To circumvent this issue, we employed an off-the-shelf Adobe Podcast toolkit\footnote{\url{https://podcast.adobe.com/enhance}}, which processes speech recordings to enhance their quality and makes them sound as if they were recorded in a professional studio. Therefore the resulting speech is better suited for our purposes. Unfortunately, due to its proprietary nature, we do not have access to its technical details. Through our observation, however, we conjecture that it may contain a pipeline of bandwidth extension \cite{su2021bandwidth} and speech enhancement \cite{su2020hifi}. Specifically, we hypothesize that the toolkit up-samples the speech to 48kHz and leverages HiFi-GAN \cite{su2021hifigan} to generate high-quality speech. We downsample the enhanced speech to the waveform sampling rate of our MRI dataset to keep model output lengths the same. In our MRI-to-speech task, we use $0.9*y_e+0.1*y_o$ as our target waveform, where $y_e$ is the enhanced waveform and $y_o$ is the original one. Using this weighted sum yields more intelligible MRI-to-speech models than using just $y_e$, which may be due to how deep speech enhancers add irregular noise that can be smoothed to more learnable targets by adding the original, more natural waveforms.

\section{Computational Efficiency}
\label{sec:computational_efficiency}

Given the importance of computational efficiency for real-time, on-device speech synthesizers we compare the number of parameters and inference times between our model and the baselines, summarized in Table \ref{space_complexity}. GPU trials use one RTX A5000 GPU, and CPU trials use none. Like Wu {\em et al.} \cite{wu2022deep}, we report inference time as the 
mean and standard deviation of five trials, each calculating the average time to synthesize an utterance in our test. Our model is faster and uses less parameters than both intermediate-representation baselines, reinforcing the idea that directly mapping articulatory features to speech is more efficient than relying on an intermediate representations.

\begin{table}[t]
\centering
\caption{\label{space_complexity}
Average inference time and number of parameters for MRI-to-speech models. See Section \ref{sec:computational_efficiency} for details.
}
\resizebox{\linewidth}{!}{
\begin{tabular}{lccc}
\toprule
\textbf{Model} & \textbf{CPU (s)} \textcolor{ForestGreen}{$\downarrow$} & \textbf{GPU (s)} \textcolor{ForestGreen}{$\downarrow$} & \textbf{Params.} \textcolor{ForestGreen}{$\downarrow$}\\
\midrule
CBL (Spe.) \cite{yu2021reconstructing} & $.66 \pm .05$ & $.081 \pm .009$ & {$1.9*10^7$} \\CBL (Hub.) & $.69 \pm .04$ & $.090 \pm .016$ & {$2.3*10^7$} \\
HiFi-CAR & $\mathbf{.58 \pm .03}$ & $\mathbf{.061 \pm .015}$ & $\mathbf{1.5*10^7}$ \\
\bottomrule
\end{tabular}}
\end{table}

\section{Synthesis Quality}
\label{sec:synthesis_quality}

\subsection{Subjective Fidelity Evaluation}
\label{sec:ab}

We perform a subjective AB preference test on Amazon Mechanical Turk (MTurk). In this evaluation, each participant is asked to distinguish between the utterances generated by our method and the baselines in terms of naturalness. We compared our model with two baselines: (1) Yu {\em et al.} \cite{yu2021reconstructing}, detailed in Section \ref{sec:cnnblstm}, and (2) \cite{yu2021reconstructing} trained with our denoised waveforms described in Section \ref{sec:enhance} as targets. For each of the two AB tests, we asked 6 native English listeners to rank a total of 9 random samples from the test set in this study. To prevent listeners from randomly submitting results, we added an audio pair consisting of one audio sample consisting entirely of noise and another audio sample containing high-fidelity speech. In Table \ref{tab:ab}, we summarize the total number of votes for each of the three options for both AB tests. Our model outperforms both baselines, receiving the most votes. The almost unanimous vote for our model in the AB test with the non-denoised baseline highlights the importance of denoising waveforms accompanying MRI data. Our model also noticeably outperforms the denoised baseline, suggesting that direct synthesis approach described in Section \ref{sec:time_domain_hifi} is well-suited for articulatory synthesis.

\begin{table}[t]
\centering
\caption{\label{tab:ab}
AB test results. See Section \ref{sec:ab} for details.
}
\begin{tabular}{lccc}
\toprule
\multirow{2}{*}{\textbf{Baseline Type}} & \multicolumn{3}{c}{\textbf{AB Test Votes}} \\ \cmidrule{2-4}
{} & \textbf{Baseline} & \textbf{Ours} & \textbf{Same}\\
\midrule
CBL (Spe.) \cite{yu2021reconstructing} & {$1$} & {$\mathbf{53}$} & {$0$} \\
+ Denoising & {$18$} & {$\mathbf{33}$} & {$3$} \\
\bottomrule
\end{tabular}
\end{table}

\subsection{Objective Fidelity Evaluation}
\label{sec:mcd}

We perform an objective evaluation of synthesis quality by analyzing the mel-cepstral distortions (MCD) \cite{black2019cmu_wilderness} between ground truths and synthesized samples, as in Wu {\em et al.} \cite{wu2022deep}. Table \ref{tab:mcd} summarizes these results, reporting the mean and standard deviation of the MCDs across utterances. Our HiFi-CAR approach outperforms both intermediate-representation baselines, suggesting that our direct modelling method is suitable for the MRI-to-speech task.

\begin{table}[t]
\centering
\caption{\label{tab:mcd}
MCD between MRI-to-speech model outputs and denoised ground truths. See Section \ref{sec:mcd} for details.
}
\begin{tabular}{lc}
\toprule
\textbf{Model} & {\textbf{MCD} \textcolor{ForestGreen}{$\downarrow$}} \\
\midrule
CBL (Spe.) \cite{yu2021reconstructing} + Denoising & {$7.31 \pm 0.45$} \\
CBL (Hub.) + Denoising & {$8.84 \pm 1.00$} \\
HiFi-CAR & $\mathbf{6.64 \pm 0.64}$ \\
\bottomrule
\end{tabular}
\end{table}

\subsection{Transcription}
\label{sec:transcription}

We also compare our method with the baselines in terms of speech intelligibility. Specifically, we compute the character error rate (CER) for speech transcription. We use Whisper \cite{radford2022robust}, a state-of-the-art automatic speech recognition (ASR) model, to generate text from the speech synthesized using each method and all test set utterances. Table \ref{tab:asr} summarizes these results. Like in Table \ref{tab:mcd}, our model outperforms both baselines, reinforcing the suitability of our model for MRI-to-speech synthesis.

\begin{table}[t]
\centering
\caption{\label{tab:asr}
ASR CER for MRI-to-speech model outputs. See Section \ref{sec:transcription} for details.
}
\begin{tabular}{lc}
\toprule
\textbf{Model} & {\textbf{CER} \textcolor{ForestGreen}{$\downarrow$}} \\
\midrule
CBL (Spe.) \cite{yu2021reconstructing} + Denoising & {$84.7\% \pm 36.4\%$} \\
CBL (Hub.) + Denoising & {$84.2\% \pm 15.7\%$} \\
HiFi-CAR & $\mathbf{69.2\% \pm 28.1\%}$ \\
\bottomrule
\end{tabular}
\end{table}

\section{Comparing MRI and EMA Features}
\label{sec:ema}

As mentioned in Section \ref{sec:intro}, MRI provides much more information about the vocal tract than EMA. MRI is a superset of EMA. Specifically, EMA has one x-y coordinate for each of the following locations: upper lip, lower lip, lower incisor, tongue tip, tongue body, and tongue dorsum. Points at all of these locations are present in the MRI data, so we can actually approximate EMA features from MRI by choosing one MRI point at each EMA location, as visualized in Figure \ref{fig:mri_apa}. In this figure, segments are the connected MRI points and the shaped dots are the estimated EMA locations. We compare these two articulatory feature sets by comparing the outputs of our proposed MRI-to-speech model with those of this model trained to synthesize speech from our estimated 12-dimensional EMA features. The test set predictions of this EMA-to-speech model yielded an MCD of $6.986 \pm 0.587$ and ASR CER of $73.2\% \pm 6.7\%$. Both of these values are worse than those of our MRI-to-speech model, summarized in Tables \ref{tab:mcd} and \ref{tab:asr}. This suggests that MRI features are more complete representations of the human vocal tract than EMA features. Thus, articulatory synthesis models should incorporate features beyond EMA in order to achieve human-like fidelity across all utterances, with MRI features being a potential feature set to extend towards. We identify which of the MRI features would be the most valuable to add to the articulatory feature set in Section \ref{sec:importance}.

\section{Identifying Important MRI Features}
\label{sec:importance}

\begin{figure}[t]
\centering
  \includegraphics[width=60mm]{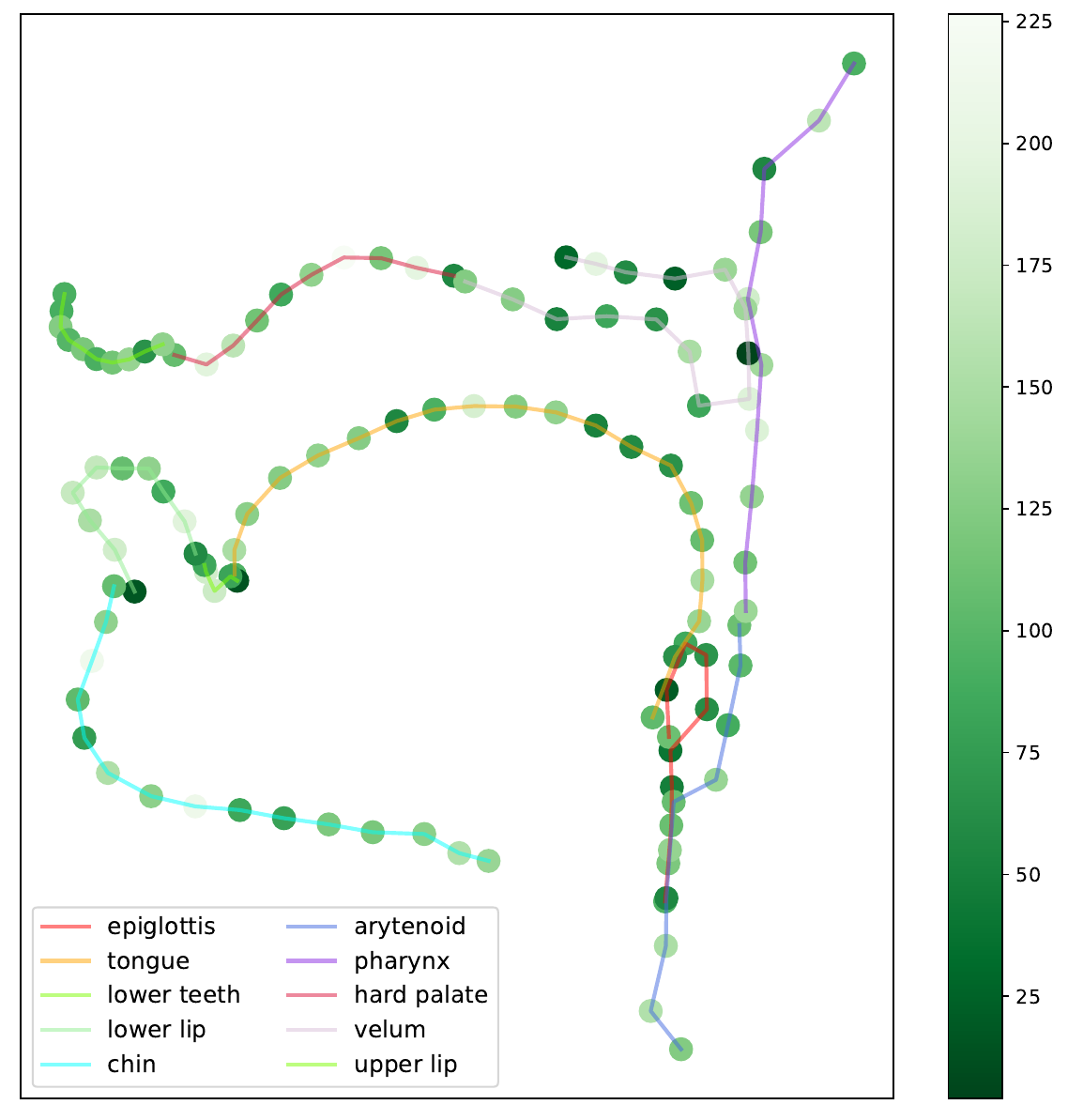}
  \caption{Importance of each MRI feature for MRI-to-speech synthesis. See Section \ref{sec:ema} for details.}
  \label{fig:mri_importance}
\end{figure}

We also study which of the MRI features are the most useful for synthesis in order to provide insight into which features should be present in an ideal articulatory feature set for articulatory synthesis. Specifically, we created 50 subsets of our 230-dimensional MRI feature set, each composed of a random 90\% subset of the 230 features. With each feature subset, we masked the 23 MRI features not in the subset to 0.0 and synthesized the test set utterances. Then, we computed the average MCD between the test set ground truths and the synthesized waveforms. For each MRI feature, we assign it a score equal to the average of the MCD values corresponding to experiments where that feature was unmasked. Since our subsets are chosen randomly, we try each feature an equal number of times in expectation. We rank the MRI features by score, with lower rank values being better and corresponding to a lower score and average MCD. Figure \ref{fig:mri_importance}, with darker green points corresponding to better MRI features. We note that each of the six EMA locations described in Section \ref{sec:ema} have an MRI point that is ranked as important. Moreover, for these six locations, besides, the upper lip, the number of points ranked as important is fairly sparse. This suggests that six points chosen in the EMA feature set are all very valuable for articulatory synthesis. The important MRI features also correspond well to the phonetic constriction task variables, {\em e.g.}, those used in \cite{sorensen2016characterizing} to model articulatory synergies from real-time MRI images. Beyond the corresponding EMA locations, points around and in the pharyngeal region ({\em e.g.}, between tongue root or epiglottis and rear pharyngeal wall) and velic region are also ranked as important. This suggests that these features are also essential for fully-specified, high-fidelity articulatory synthesis. The pharyngeal features are relevant to the production of various speech sounds \cite{moisik2019putting}, like /a/ \cite {proctor2015articulation} and some variants of /r/ in English \cite{alwan1997toward}. The velic features are crucial to the production of nasal sounds. Both of these features are not available from EMA. Thus, moving forward, we plan to incorporate pharyngeal and velic features in all of our articulatory synthesis models. Points around constriction locations, whether at the lips, tongue, or throat, are generally ranked as important. Thus, when designing sparse articulatory feature sets, it may be useful to prioritize these constriction locations.

\section{Conclusion and Future Directions}

In this work, we devise a new articulatory synthesis method using MRI-based features, providing preprocessing and modelling strategies for working with such data. Based on MCD, ASR, human evaluation, timing, and memory measurements, our model achieves noticeably better fidelity and computational efficiency than the prior intermediate-representation approach. Through speech synthesis ablations, we also show the advantages of MRI over EMA and identify the most important MRI features for articulatory synthesis. Moving forward, we will extend our work to multi-speaker synthesis and inversion tasks \cite{lim2021multispeaker, wu2023speaker}.

\bibliographystyle{IEEEtran}
\bibliography{mybib}

\end{document}